\documentstyle[prl,aps,epsf]{revtex}
\begin{document}
\draft
\twocolumn[\hsize\textwidth\columnwidth\hsize\csname @twocolumnfalse\endcsname
\title{Can liquid metal surfaces have hexatic order~?}
\author{
Franck Celestini$^{1,2}$,
Furio Ercolessi$^{1,2}$, and
Erio Tosatti$^{1,2,3}$
}
\address{
$^{\rm (1)}$ International School for Advanced Studies (SISSA-ISAS),
I-34014 Trieste, Italy}
\address{
$^{\rm (2)}$ Istituto Nazionale di Fisica della Materia (INFM), Italy}
\address{
$^{\rm (3)}$ International Centre for Theoretical Physics (ICTP),
I-34014 Trieste, Italy}
\date{21 December 1996}
\maketitle
\begin{center}
Preprint SISSA {\bf 186/96/CM/SS} \\
{\sl Physical Review Letters}, 1997
\end{center}
\begin{abstract}
We propose that extended orientational correlations can
appear at the surface of supercooled heavy noble
liquid metals, due to the same compressive forces
that cause reconstruction of their crystal surfaces.
Simulations for liquid  Au show a packed surface layer
structurally akin to a defected 2D triangular solid.
Upon supercooling, the density of unbound disclinations decreases as expected
in Nelson-Halperin's theory. It extrapolates to a hexatic transition
about 350 K below melting, where a sharp growth of orientational correlation
length and time is also found. True hexatic order 
is preempted by surface-initiated recrystallisation.

\end{abstract}
\vskip2pc]

\narrowtext

Heavy noble metals  such as Au, Pt or Ir have a general tendency to reduce 
their lateral interatomic distance at the surface, relative to the bulk. This
tendency, whose microscopic origin has been traced to a relativistic 
intra-atomic s-d energy shift \cite{takeuchi}, accounts for 
the zero-temperature reconstructions of the crystal surfaces. Remarkably, 
the lateral surface compression does not heal out with temperature. On
the contrary, it is seen to persist, and in fact 
to increase, as temperature rises. For example, the lateral 
first-layer atomic density of Au(111) and Pt(111) has been found to 
increase, relative to the bulk layer density, from 1.04 to 1.07, and from 
1.0 to 1.05 respectively, between room temperature and 1000 K \cite{Sandy}.  
Our present rationale for this somewhat surprising behavior is that at high 
temperatures the substrate periodic potential felt by the first
layer atoms is progressively reduced by thermal fluctuations and by a large
outwards expansion, so that shrinkage is less hindered by epitaxy.

In the liquid metal surface, which constitutes our present concern,
the periodic potential is absent altogether. High 
lateral densities, now unhindered, can be expected to persist, 
in conjunction with strong surface layering, at least for temperatures 
which are not too high. Close to the melting 
point, X-ray data have revealed
very pronounced layering oscillations in the density profiles normal to 
the surface in liquid metals such as Hg \cite{rice} and Ga
\cite{magnussen}, implying first layer lateral densities larger than
the bulk average. Based on simulations, we pointed out some time ago 
\cite{iarlori} that still larger layering oscillations, and a higher
surface lateral density, should be expected for the heavy
noble liquid metal surfaces, boosted by the compressive
tendencies just discussed. In addition, we found that supercooling makes
the effect even stronger, with a nearly 2D-crystalline surface layer.
This naturally raised the question, which we resolve in this Letter,
whether it could be possible for hexatic, orientational ordering, or at 
least for very extended orientational correlations, to appear at
the free surface of a heavy noble liquid metal, such as Au, Pt,
or Ir, under suitable supercooling conditions.

The original suggestion made by Nelson and
Halperin\cite{Nelson79}
(extending earlier work of Kosterlitz and Thouless \cite{KT}),
and pursued by Young \cite{young} and others \cite{StrandBook},
that in 2D an intermediate hexatic phase could appear  between the solid and 
the liquid, led to a wide search in a variety of 
systems \cite{StrandBook}. Beautiful realizations have been achieved 
such as those in 2D colloidal systems
\cite{murray}, magnetic bubble lattices \cite{seshadri},
and vortex lattices \cite{grier}.
On the other hand, the 2D simulation studies have 
been exceedingly controversial \cite{StrandBook}. An important clue 
which emerged more recently appears to be the choice of thermodynamic
ensemble, with a strong preference of hexatic phenomena to show up in
ensembles where density changes are naturally allowed
\cite{bagchi96}. In grand-canonical conditions, the binding of disclinations 
taking place at the fluid-hexatic transition is facilitated, since it
does not rely on the slow, 
canonical diffusion of defects. For instance, spontaneous creation 
of a new disclination near a pre-existing one, can lead to a much faster
appearance of the dislocation bound state.

In the liquid metal surface, atom exchange between layers can be extremely
effective, with typical atomic diffusion times of order 100 ps 
between first and second ``layer'' for Au  at the melting point \cite{front}. 
This free interchange
of particles with the liquid substrate makes the outermost
layer of the liquid metal a 2D system which is naturally
``grand-canonical'' on a very fast, nanosecond time scale. This 
is fast enough, in particular,  that even ordinary, canonical Molecular 
Dynamics (MD) simulation should be able to equilibrate orientational 
correlations without problems. 

Moved by these motivations, we conducted very extensive MD simulations of the
liquid gold surface. We used a slab geometry with two free 
surfaces, a thickness of about $\simeq\rm 60\,\AA$ and a lateral size of
$\simeq\rm 120\,\AA$, for a total of 50000 atoms. A second, smaller 
system with the same thickness but a lateral size of $\simeq\rm 80\,\AA$ 
was also studied in order to monitor size effects. Periodic boundary
conditions were required parallel to the slab. Atoms interacted through 
the glue model hamiltonian which is well documented for 
this system, providing a good description of 
many properties of bulk solid 
and liquid Au, of the melting temperature, ($T_m = 1320-1350\,\rm K$,
against an experimental value of 1336), of all the main solid surfaces 
including their detailed energies and reconstructions, and of the liquid 
surface tension at the melting point\cite{philmag88,iarlori}. 
Constant-energy runs were used to generate atomic trajectories. Data 
were taken from 2000 K, well above $T_m$, 
to 950 K, well below. Data 
below $T_m$ clearly refer to a 
supercooled liquid slab, whose lifetime 
we nevertheless found to exceed 1 ns down to $T\sim 1000\,\rm K$. 
Below 1000 K, the lifetime
dropped, and it became impossible to equilibrate the liquid, due to (111)
recrystallization rapidly nucleating at both slab surfaces. Above 1000 K, 
where 
the system remained liquid, we worked at temperature 
intervals of 50
K up to $T_m$, then more sparsely.  At each temperature, 
typically 
30000 steps (one time-step $\sim$ $10^{-14}$ s) were used first for
equilibration. Then, 20 uncorrelated configurations (40 for the 
smaller system) were generated from successive 10000-step runs, 
and subsequently analysed. 

The first step of the analysis consists in
identifying the surface topmost, ``first-layer'' atoms in each configuration. 
That was done using a prescription which proved to work well in the 
past \cite{iarlori}. By representing atoms as spheres of radius
$R$, we define surface atoms those which remain fully visible, i.e., totally
unshadowed from above the slab. The optimal $R$ value, corresponding
to atoms in the first density profile peak in Fig.\ 1, 
was found to be 
$1.8\,\rm\AA$, reasonably close to the atomic radius.  Fig.\ 1 
shows  a Delaunay triangulation map \cite{delaunay} for the surface atoms,
where disclinations (five- and seven-fold) are pinpointed.
At high temperature there is a high concentration 
of defects, both bound and free, while the density profile surface peak is
relatively weak. Conversely,  the peak is 
very strong at low temperatures, where the map 
shows large triangular domains, and a  reduced density of free
disclinations $n_f$. 
Theory predicts \cite{Nelson79} a behavior of the form
\begin{equation}
n_f \simeq \exp\left[
\frac{-2a}{(T-T_i)^{1/2}}
\right]
\label{eq:nf}
\end{equation}
as the fluid-hexatic transition temperature $T_i$ is approached. We 
find that this equation fits very closely our simulation data, 
with  $a\simeq 2.2\,\rm K^{1/2}$ and $T_i\simeq 974\,\rm K$
(Fig.\ 2).
This suggests that the liquid Au surface would
\begin{figure}
\vspace*{0.5cm}
\epsfxsize=7cm
\epsfbox{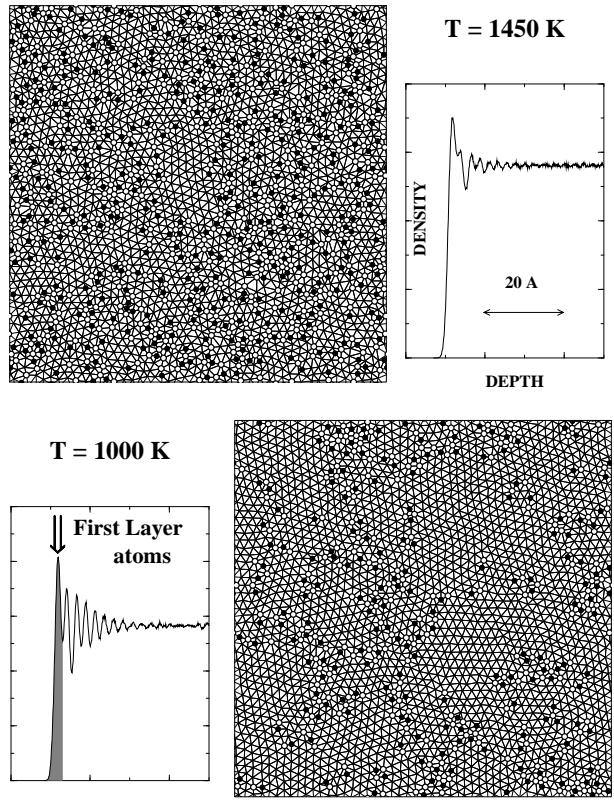}
\vspace*{0.7cm}
\caption{
Delaunay triangulation maps for the surface atoms of simulated liquid Au
(larger system) at $T=1450\,\rm K$ (above $T_m$) and $T=1000\,\rm K$
(supercooled).  
Fivefold (black) and sevenfold (white) disclinations are pinpointed. 
The two corresponding density profiles along the surface 
normal are also shown. 
The first peak correspond to first layer atoms in the maps.
}
\label{fig:1}
\end{figure}
\begin{figure}
\vspace*{-1.2cm}
\epsfxsize=9cm
\epsfbox{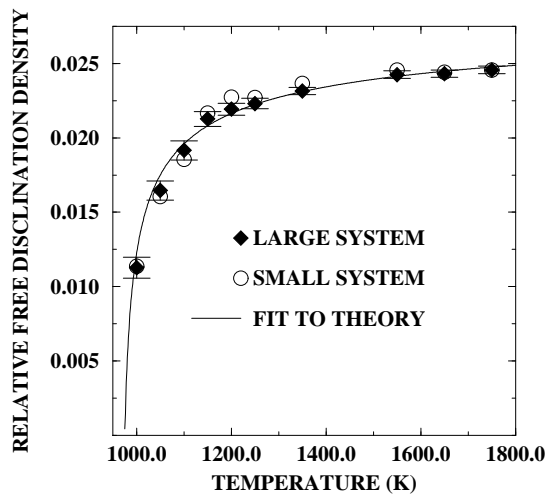}
\vspace*{-4.2cm}
\caption{
Density of free disclinations (of both signs) for the liquid metal surface 
as a function of temperature. Full diamonds: larger system; open circles: 
smaller system; Solid curve: fit to Eq.~(\protect\ref{eq:nf}).
}
\label{fig:2}
\end{figure}
\noindent
undergo, in its supercooled metastable state, a genuine
fluid-hexatic transition, if one could somehow prevent 
recrystallization from taking place.

Encouraged by this result, we turned to a comparison of positional and 
orientational spatial correlations, and of their dynamics, versus 
temperature. The static orientational correlation function is defined as
\begin{equation}
g_6 (r,t=0) = \langle O_6^{i*}(r,t=0) O_6^j (0,t=0) \rangle
\end{equation}
where the orientational order parameter $O_6$ 
is, at surface site $i$  
\begin{equation}
O_6^i = \frac{1}{N_i} \sum_{j=1}^{N_i} \exp(6i\theta_{ij}) .
\end{equation}
The sum is restricted to the $N_i$ Delaunay surface neighbors of atom $i$, 
and $\theta_{ij}$ is the angle formed by the $ij$ bond (projected on the $xy$
plane) with 
the $x$ axis. Positional correlations are defined similarly, using
the surface atom local 2D density as the order parameter. Correlation 
lengths $\xi_p$ and $\xi_6$ are obtained by fitting the envelope of 
the computed correlation functions to an exponential decay, 
$g(r)-1$ $\sim$ $\exp(-r/\xi_p)$. Fig.\ 3a  shows that whereas
above 1200 K
positional and orientational surface 
correlation lengths are close, and about
 $5\,\rm\AA$, below this temperature the latter take off while the former
do not. A fit of $\xi_6(T)$ to theory, which predicts 
$\xi_6(T) \simeq \exp\left[ a/(T-T_i)^{1/2} \right]$ yields 
reasonable agreement for our larger system, with $a\simeq 15.5\,
\rm K^{1/2}$ and $T_i\simeq 942\,\rm K$. There is however an
alarmingly large discrepancy with the parameter values of the
disclination fit. The disagreement turns out to be instructive, 
and is explained by a size effect. By comparing our small and large
systems, we observe that whereas disclination densities hardly 
depend on size, the growth of long-ranged orientational correlations 
is enormously hampered by size. As Fig.\ 3a shows, a decrease of lateral
size from 120 to $80\,\rm\AA$ is sufficient to damage the rise of   
$\xi_6$, reducing it (at 1000 K) from 12.5 to $7.5\,\rm\AA$. 
This is, we surmise, an independent signature of the Coulomb nature
of effective interactions between disclinations in the liquid metal
surface. Size effects in two-body correlations are known to be extremely 
large in related systems such as the XY model \cite{binder}.
Fortunately (a proper size scaling being
presently out of the question), the one-body disclination density
is instead well converged (Fig.\ 1). We conclude
that, even though $\xi_6$ is still far from
convergence even in the large system, the disclination fit 
parameters should safely represent the infinite system.   

As theory suggests \cite{zippelius}, and as the colloidal experiments have 
demonstrated \cite{murray}, the
temperature-dependent orientational relaxation times can provide 
an even stronger dynamical signature of hexatic
phenomena. We  computed  the positional and orientational
correlation times $\tau_p$ and $\tau_6$ by an exponential time fit of the 
envelope
\begin{figure}
\vspace*{-1.1cm}
\epsfxsize=10cm
\epsfbox{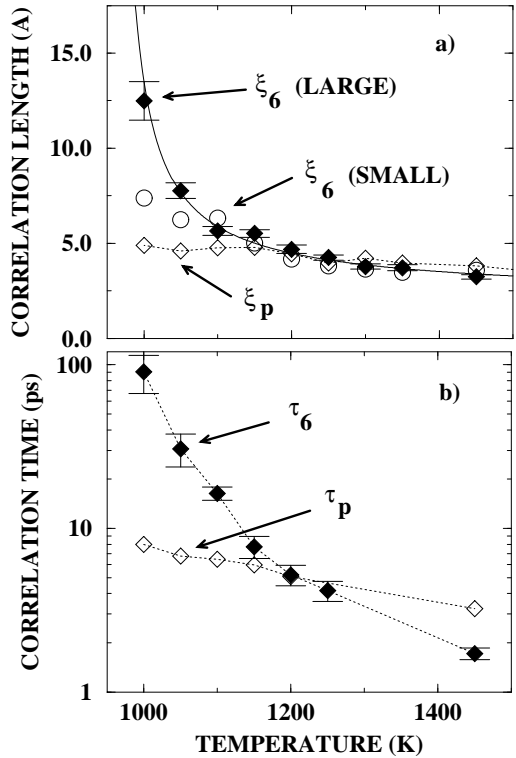}
\vspace*{-2cm}
\caption{
(a) Orientational (full diamonds) and positional (empty diamonds)
correlation lengths. Open circles: orientational length for the smaller
system. The full line is a fit to theory (see text).
(b) Orientational (full diamonds) 
and positional (empty diamonds) correlation times. Dotted lines are guides
to the eyes.
}
\label{fig:3}
\end{figure}
\noindent
of the time-dependent 2D structure factor $F(K,t)$,
where $K$ is the frustrated reciprocal lattice spacing,
and of the time-dependent orientational correlation function
$g_6 (r=0,t)$. Fig.\ 3b
shows the results for the larger system. Below about 1200 K 
$\tau_6$ is found to rise dramatically (note the log scale) 
 while $\tau_p$ does not. There is a clear orientational
slowing down, which complements very well the static results, and also 
suggests that a likely experimental signature  to be sought at 
the liquid metal surface might be dynamical rather than static.

In spite of these growing correlations, true hexatic long-range order
appears, in our particular system, to be narrowly preempted by
recrystallization.  Already at 1000 K, after a lifetime of
about 0.5 ns the liquid slab suddenly starts recrystallizing. 
At 950 K, the liquid lifetime (starting with the 1000 K liquid
configuration) drops to less than 0.1 ns. Recrystallization proceeds
with two fronts starting simultaneously at the two surfaces. 
At 1000K, we analysed first the metastable liquid surface,
and then the moving recrystallization front,
monitoring values of $\xi_6$ and $\xi_p$ in the successive layers as a 
function 
of depth \cite{front}. As shown in fig.\ 4 
the front is split, an orientational ordering front 
{\it preceding} a positional ordering, 
\begin{figure}
\vspace*{-2cm}
\epsfxsize=10cm
\epsfbox{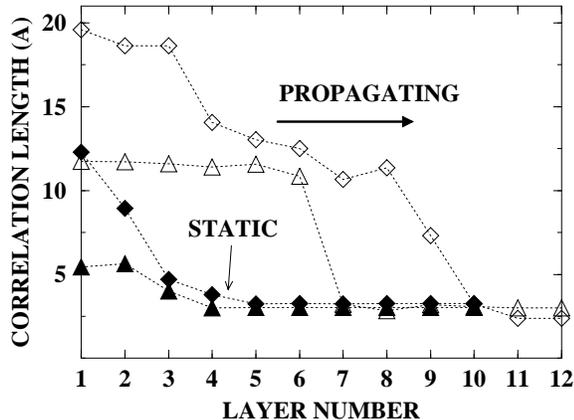}
\vspace*{-5cm}
\caption{
Orientational (diamonds) and positional (triangles) surface depth profiles.
Bottom: liquid surface, T=1000 K. Top: during recrystallization. Note that
the orientational front precedes the positional front.
}
\label{fig:4}
\end{figure}
\noindent
crystalline front. The intermediate 
region, where 
orientational correlations are extended, but positional
correlations are short, is strongly layered (not shown), and extends for a 
thickness of about three layers. In fact layers 
just below the surface already possess, in the purely liquid slab, long
orientational correlations (fig.\ 4). This kind of 
surface wetting by an orientationally ordered film coexists,
as we verified directly, with rough, liquid-like height-height 
static correlations (not shown), and 
seems an exclusive characteristics of a strongly layered liquid surface.
No such orientationally ordered film is found, for example, in a 
Lennard-Jones system, either at the 
 free liquid surface, or at the crystal-liquid interface
\cite{tallon}. At the liquid metal surface, the presence of the 
orientationally ordered film seems to provide the required germ for
propagation of the crystal phase into the bulk. 

In conclusion, the supercooled surface of liquid Au, and most likely 
also that of other heavy noble metals such as Pt or Ir, is predicted 
to develop an incipient long-ranged orientational order described
by the Nelson-Halperin theory. The order
extends a few layers below the surface. Transition to a well defined
hexatic phase is preempted by a fast  and effective surface-driven 
recrystallisation, where the orientationally ordered film precedes
the crystalline front. Our results should motivate future 
experiments, either structural, to detect the orientationally 
quasi-ordered film, or dynamical, addressing the orientational slowing
down, or metallurgical, to investigate the nature and depinning of the
recrystallization front in these simple but exciting systems.

~

~

~

~

\begin{sloppypar}
It is a pleasure to thank S. Iarlori, G. Santoro and A. Parola
for illuminating discussions.
We acknowledge partial support from the European Commission under contract 
ERBCHRXCT920062 and ERBCHRXCT930342. Work at SISSA by F.\ C.\ is 
partly under European Commission
sponsorship, contract ERBCHBGCT940636.
\end{sloppypar}




\begin{references}
\bibitem{takeuchi}  N. Takeuchi, C. T. Chan, and K. M. Ho,
  Phys. Rev. Lett. {\bf 63}, 1273 (1989); J. F. Annett and
  J. E. Inglesfield,
  J. Phys.: Cond. Matt. {\bf 1}, 3645 (1989);  V. Fiorentini,
  M. Methfessel, and M. Scheffler,
  Phys. Rev. Lett. {\bf 71}, 1051 (1993).
  
\bibitem{Sandy}
A. R. Sandy, S. G. J. Mochrie, D. M. Zehner, G. Grubel, K. G. Huang and 
Doon Gibbs,
Phys. Rev. Lett. {\bf 68}, 2192 (1992).

\bibitem{rice}
B. C. Lu and S. A. Rice, 
J. Chem. Phys. {\bf 68}, 5558 (1978).  

\bibitem{magnussen}
M. J. Regan, E. H. Kawamoto, S. Lee, P. S. Pershan, N. Maskil, M. Deutsch,
O. M. Magnussen, B. M. Ocko and L. E. Berman,
Phys. Rev. Lett. {\bf 75}, 2498 (1995).

\bibitem{iarlori}
S. Iarlori, P. Carnevali, F. Ercolessi, and E. Tosatti,
Surf. Sci. {\bf 211/212}, 55 (1989);   
Europhys. Letters {\bf 10}, 329 (1989).

\bibitem{Nelson79}
  D. R. Nelson and B. I. Halperin,
  Phys. Rev. B {\bf 19}, 2457 (1979).
  
\bibitem{KT}
J. M. Kosterlitz and D. J. Thouless,
J. Phys. C {\bf 6}, 1181 (1973).

\bibitem{young}
A. P. Young, Phys. Rev. B {\bf 19}, 1855 (1979).
  
\bibitem{StrandBook}
See, e.g., {\em Bond-orientational order in condensed matter systems}, 
K. Strandburg (Ed.), Springer, New York, 1992;
and references therein.  

\bibitem{murray} 
C. A. Murray and R. A. Wenk,
Phys. Rev. Lett. {\bf 62}, 1643 (1989);
C. A. Murray, W. O. Sprenger, and R. A. Wenk,
Phys. Rev. B {\bf 42}, 688 (1990).

\bibitem{seshadri}
R. Seshadri and R. M. Westervelt,
Phys. Rev. Lett. {\bf 66}, 2774 (1991).

\bibitem{grier}
D. G. Grier {\em et al.},
Phys. Rev. Lett. {\bf 66}, 2270 (1991).

\bibitem{bagchi96} 
K. Bagchi and H. C. Andersen,
Phys. Rev. Lett. {\bf 76}, 255 (1996);
K. Chen, T. Kaplan, and M. Mostoller,
Phys. Rev. Lett. {\bf 74}, 4019 (1995).

\bibitem{philmag88}
F. Ercolessi, M. Parrinello and E. Tosatti 
  Phil.\ Mag.\ A{\bf 38}, 213 (1988);
  E. Tosatti and F. Ercolessi,
  Mod.\ Phys.\ Lett.\ B {\bf 5}, 413 (1991).
  
\bibitem{delaunay}
F. F. Preparata and M. L. Shamos,
  {\em Computational Geometry : An introduction}, 
 Springer, New York, (1985).

\bibitem{binder}
For references see, e.g.,
{\em Monte Carlo methods}, 
K. Binder (Ed.), Springer, Berlin, 1987.  

\bibitem{zippelius}
A. Zippelius, B. I. Halperin and D.R. Nelson,
Phys. Rev. B {\bf 22}, 2514 (1980).

\bibitem{front}
  F. Celestini, F. Ercolessi, and E. Tosatti
(unpublished).
  
\bibitem{tallon}
  J. Tallon,  
  Phys. Rev. Lett. {\bf 57}, 1328 (1986).  
    
\end{references}
\end{document}